\documentclass[acmsmall]{acmart}
\renewcommand\footnotetextcopyrightpermission[1]{} 
\usepackage[utf8]{inputenc}
\usepackage{url}
\expandafter\def\expandafter\UrlBreaks\expandafter{\UrlBreaks
  \do\a\do\b\do\c\do\d\do\e\do\f\do\g\do\h\do\i\do\j%
  \do\k\do\l\do\m\do\n\do\o\do\p\do\q\do\r\do\s\do\t%
  \do\u\do\v\do\w\do\x\do\y\do\z\do\A\do\B\do\C\do\D%
  \do\E\do\F\do\G\do\H\do\I\do\J\do\K\do\L\do\M\do\N%
  \do\O\do\P\do\Q\do\R\do\S\do\T\do\U\do\V\do\W\do\X%
  \do\Y\do\Z}
\usepackage{booktabs}

\usepackage{verbatim}
\usepackage{subfig}
\usepackage{xcolor}
\usepackage{soul}

\usepackage{amsthm}
\usepackage{algorithm}
\usepackage{multirow}
\usepackage{algorithmic}

\usepackage{graphicx}
\usepackage{epsf}
\usepackage{tikz}
\usepackage{tikzscale}
\usepackage{pgfplots}
\usepgfplotslibrary{external}
\usetikzlibrary{external}
\usepackage{amsmath}
\DeclareMathOperator{\fingerprint}{fingerprint}

\DeclareMathOperator{\map}{map}
\DeclareMathOperator{\assign}{assign}

\newcommand{\stack}{\ensuremath{\sigma}}

\DeclareMathOperator{\xor}{\mathrm{~xor~}}
\pgfplotsset{compat=1.14}

\usetikzlibrary{positioning}

\setcopyright{acmlicensed}
\acmJournal{JEA}
\acmYear{2019} \acmVolume{1} \acmNumber{1} \acmArticle{1} \acmMonth{1} \acmPrice{15.00}\acmDOI{10.1145/3376122}

\begin{document}

%
\title{Xor Filters: Faster and Smaller Than Bloom and Cuckoo Filters}

%
\author{Thomas Mueller Graf}
\email{thomas.tom.mueller@gmail.com}
\author{Daniel Lemire}
\orcid{1234-5678-9012}
\authornotemark[1]
\email{lemire@gmail.com}
\affiliation{%
  \institution{University of Quebec (TELUQ)}
  \streetaddress{5800 Saint-Denis, Office 1105}
  \city{Montreal}
  \state{Quebec}
  \country{Canada}
  \postcode{H2S 3L5}
}

\begin{abstract}
The Bloom filter provides fast approximate set membership while using little memory.
Engineers often use these filters to avoid slow operations such as disk or network accesses. As an alternative, a cuckoo filter may need less space than a Bloom filter and it is  faster. Chazelle et al. proposed a generalization of the Bloom filter called the Bloomier filter. Dietzfelbinger and Pagh described a variation on the Bloomier filter that can answer approximate membership queries over immutable sets. It has never been tested empirically, to our knowledge. We review an efficient implementation of their approach, which we call the xor filter. We find that xor filters can be faster than Bloom and cuckoo filters while using less memory. We further show that a more compact version of xor filters (xor+) can use even less space than highly compact alternatives (e.g., Golomb-compressed sequences) while  providing speeds competitive with Bloom filters.
\end{abstract}

\ccsdesc[500]{Theory of computation~Bloom filters and hashing}

\keywords{Bloom Filters, Cuckoo Filters, Approximate Set Membership}
\maketitle 
\thispagestyle{empty}
\section{Introduction}



The classical data structure for approximate membership is the Bloom filter~\cite{Bloom:1970:STH:362686.362692}. It may be the best-known probabilistic data structure. A Bloom filter is akin to a set data structure in that we can add keys, and  check whether a given key is present in the set. There is a small probability that a key is incorrectly reported as being present, an event we call a \emph{false positive}. However, Bloom filters can use less memory than the original set. Thus, Bloom filters accept a small probability of error for a reduced memory usage.

Approximate set membership has many applications: e.g., scanning for viruses using payload signatures~\cite{erdogan2007hash}, filtering bad keywords or addresses, and fast language identification for strings~\cite{Jacob:2007:LCU:1328554.1328564}.
Write-optimized key-value stores~\cite{Chang:2008:BDS:1365815.1365816}
such as log-structured merge (LSM) trees~\cite{o1996log} are another important use case. 
In such stores, 
 an in-memory data structure 
avoids expensive disk accesses. 

We want our data structures to be fast and to use little memory.
In this respect, conventional Bloom filters can be surpassed:
\begin{itemize}
\item Bloom filters generate many random-access queries.
For efficient memory usage, a Bloom filter with  a false-positive probability $\epsilon$ should use about $-\log_2 \epsilon$~hash functions~\cite{broder2004network}. At a false-positive probability of 1\%, seven hash functions are thus required. Even if the computation of the hash functions were free, doing many random memory accesses can be expensive. 


\item The theoretical lower bound for an approximate membership data structure with a false-positive probability $\epsilon$ is $-\log_2 \epsilon$~bits per key~\cite{broder2004network}. When applied in an optimal manner, Bloom filters use 44\% more memory than the theoretical lower bound.
\end{itemize}

Practically, Bloom filters are often slower and larger than alternatives such as cuckoo filters~\cite{Fan:2014:CFP:2674005.2674994}. 
Can we do better than even cuckoo filters?

Bonomi et al.~\cite{Bonomi:2006:ICC:1276191.1276252} as well as Broder and Mitzenmacher~\cite{broder2004network} remarked that
for static sets, essentially optimal memory usage is possible using a \emph{perfect hash function} and fingerprints. 
They dismissed this possibility in part 
because perfect hash functions might be too expensive to compute. 
Yet Dietzfelbinger and Pagh~\cite{10.1007/978-3-540-70575-8_32} described a seemingly practical implementation
of this idea which we call an xor filter. It builds on closely related work such as Bloomier filters~\cite{Chazelle:2004:BFE:982792.982797, charles2008bloomier}. 

To our knowledge, xor filters were never implemented and benchmarked. %
We present the first experimental evaluation. 
We find that they perform well, being often faster than both Bloom and cuckoo filters.
For common use cases, they require less memory.
Furthermore, we can  improve their memory usage with only a modest performance penalty, using
a relatively simple compression technique (see \S~\ref{sec:xorplus}).
We make our software freely available to ensure reproducibility.

Our main result is that xor filters have merit as a practical data structure. They are fast, compact
and we found them easy to implement. 

\section{Related Work}

We find many Bloom filters and related data structures  within database systems~\cite{Chang:2008:BDS:1365815.1365816} to avoid disk accesses. A popular strategy for designing database engines that must support frequent updates is the log-structured merge (LSM) tree~\cite{o1996log}. At a high-level, LSM trees maintain  a fast in-memory component that is merged, in batches, to data in persistent storage. The in-memory component accumulates database updates thus amortizing the update cost to persistent storage. 
To accelerate lookups, many LSM tree implementations (e.g., levelDB,  RocksDB,  WiredTiger) use Bloom filters. 
When merging the components, usually a new filter is built.
We could, instead, update existing filters. However, data structures that support fast merging (e.g., Bloom filters)
require either the original filters to have extra capacity, or
the result of the merger to have higher false-positive probabilities~\cite{almeida2007scalable}.

Many applications of Bloom filters and related data structures are found in networking, where we seek to avoid unnecessary network access.
Generally, whenever a filter must be sent through a network connection to other computers (e.g., to cache and prevent network queries), we might be able to consider the filter as immutable~\cite{Mitzenmacher:2002:CBF:581876.581878} on the receiving machine.

\subsection{Bloom Filter Variants}
\emph{Standard Bloom filters}~\cite{Bloom:1970:STH:362686.362692} consist of a collection of  hash functions $h_1$, $h_2$, \ldots, $h_k$, which map each possible key to a fixed integer which we interpret as an index value, and an array of bits $B$, initialized with zeros. The size of the array and the number of hash functions $k$ are parameters of the filter.
When we add a key $x$, we hash it with each hash function, and set the corresponding bits: \begin{eqnarray*}B[{h_1(x)}] & \leftarrow 1,\\ B[{h_2(x)}] & \leftarrow 1,\\ & \vdots \\ B[{h_k(x)}] & \leftarrow 1.\end{eqnarray*} To determine whether a given key is likely present, we check that the corresponding bits in our array are set: \begin{eqnarray*}(B[{h_1(x)}] = 1) \mathrm{~and~} (B[{h_2(x)}] = 1) \mathrm{~and~} \cdots \mathrm{~and~} (B[{h_k(x)}] = 1).\end{eqnarray*}
 Thus, if there are $k$~hash functions, we might need to check up to $k$~bits. For keys that were added, we are guaranteed that all bits are set: there can never be a false negative. But false positives are possible, if the bits were set by other keys. The standard Bloom filter does not allow us to remove keys.
Bloom filters  support adding keys irrespective of the size of the bit array and of
the number of hash functions, but the false-positive probability increases as more entries are added, and so more bits are set.

The size of the array $B$ is typically chosen so that a certain false-positive probability can be guaranteed up to a maximal number of entries, and the  optimal parameter $k$ is calculated. 
The expected space overhead for optimal Bloom filters is 44\%: it requires setting $k = - \log_2 \epsilon$ where $\epsilon$ is the desired bound on the false-positive probability. 
Bloom filters can be made concurrent~\cite{5476244}.

\emph{Blocked Bloom filters}~\cite{Putze:2010:CHS:1498698.1594230,lang2019performance} consist of many small Bloom filters, maybe one per CPU cache line, so that they need only one memory access per operation. However, the load of those small filters is likely to be uneven, and so for the same false-positive probability, they often need about 30\% more space than standard Bloom filters. Advanced CPU instructions allow to speed up membership tests for both regular and blocked Bloom filters~\cite{Polychroniou:2014:VBF:2619228.2619234}.

There are many other variations on Bloom filters including counting Bloom filters~\cite{Bonomi:2006:ICC:1276191.1276252,Rottenstreich:2014:VCB:2736196.2736202} which support removing keys at the expense of more storage, compressed Bloom filters~\cite{Mitzenmacher:2002:CBF:581876.581878}, multidimensional Bloom filters~\cite{Crainiceanu:2015:BLO:2825276.2825463}, Stable Bloom filters~\cite{Deng:2006:ADD:1142473.1142477} and so forth.

\subsection{Fingerprint Based Variants}
Fingerprint-based variants store a fingerprint per key, where a fingerprint is  the result of  hash function $h$; typically, it is a word having a fixed number of bits. The membership test  consists of the retrieval and comparison  with the relevant fingerprints for the given key.
The general intuition is as follows. For each value $x$ in the set, we store the fingerprint $h(x)$ in a key-fingerprint data structure. 
Given a candidate value $y$, we access its fingerprint from the data structure and we compare the result with $h(y)$. Whenever $y$ was part of the set, the fingerprints match, otherwise they are likely different with a probability that depends on the size of the fingerprint.

\begin{itemize}
\item \emph{Golomb-compressed sequences}~\cite{Putze:2010:CHS:1498698.1594230} store the sorted fingerprints by encoding the differences between fingerprint values. The overhead of this encoding is at least 1.5~bits per key, but it is difficult to achieve competitive  speed. 

\item \emph{Cuckoo filters}~\cite{Fan:2014:CFP:2674005.2674994} are based on cuckoo hashing. At full capacity, and with a low false-positive probability, they use less space than Bloom filters, and membership tests are often faster. The overhead is 3~bits per key for the standard cuckoo filter, and 2~bits per key for the slower semi-sorted variant.  We are not aware of a cuckoo filter implementation that supports concurrent updates though there are  related cuckoo hashing concurrency strategies~\cite{Li:2014:AIF:2592798.2592820}.

\item \emph{Quotient filters}~\cite{Pandey:2017:GCF:3035918.3035963} store fingerprints in a compact hash table. Quotient filters and cuckoo filters use a similar amount of memory.

\item \emph{Morton filters}~\cite{Breslow:2018:MFF:3213880.3232248} are similar to cuckoo filters, but use underloaded buckets, like Horton tables~\cite{Breslow:2016:HTF:3026959.3026986}.
Many sparse buckets are combined into a block so that data is stored more densely. 

\item \emph{Bloomier filters}~\cite{Chazelle:2004:BFE:982792.982797, charles2008bloomier} support approximate evaluation of arbitrary functions, in addition to approximate membership queries. 
We are interested in a  variant of the Bloomier filter~\cite{10.1007/978-3-540-70575-8_32}
that can be used for approximate membership queries. We call this variant the xor filter (\S~\ref{xor}).

\end{itemize}
Other variants have been proposed~\cite{10.1007/978-3-642-03351-3_25,10.1007/978-3-319-94144-8_24} but authors sometimes omit to provide and benchmark practical implementations.
Dietzfelbinger and Pagh~\cite{10.1007/978-3-540-70575-8_32} observe that fingerprint techniques can be extended by storing auxiliary data with the fingerprint.

\section{Xor Filters}
\label{xor}


Given a key $x$, we produce its
$k$-bit fingerprint (noted $\fingerprint(x)$) using a randomly chosen hash function. We assume an idealized fully independent hash function; all fingerprints are equally likely so that  $P(\fingerprint(x) = c) =  1/2^k$ for any $x$ and $c$. 
This probability $\epsilon = 1/2^k$ determines the false-positive probability of our filter. 
We summarize our notation in Table~\ref{table:notation}. 





We want to construct a map $F$ from all possible elements to
$k$-bit integers such that it maps all keys $y$
from a set $S$ to their $k$-bit  $\fingerprint(x)$.
Thus, if we pick any element of the set, it
gets mapped to its fingerprint by design $F(y)=\fingerprint(y)$. Any value that is not
part of the filter  gets mapped to a value distinct from its fingerprint
with a probability $1-\epsilon = 1- 1/2^k$.

We store the fingerprints in an array $B$ with capacity $c$ slightly larger than
the cardinality of the set $|S|$ (i.e., $c \approx 1.23 \times |S|$).
We randomly and independently choose
three hash functions
$h_0, h_1, h_2$ from $U$ to consecutive ranges of integer
values ($h_0: S \to \{0,\ldots, c/3 - 1\} $, $h_1: S \to \{c/3,\ldots,2c/3 - 1\} $, $h_2: S \to \{2c/3,\ldots,c-1\} $).
For example, if $c=12$, we might have the ranges $\{0,\ldots, 3\}$,
$\{4,\ldots, 7\}$, and
$\{8,\ldots,11\}$.
Our goal is to have that the exclusive-or aggregate of the values in array $B$ at
the locations given by the three hash functions agree with the fingerprint ($B[{h_0(x)}] \xor B[{h_1(x)}] \xor B[{h_2(x)}] = \fingerprint(x)$) for all elements $x \in S$. The hash functions $h_0, h_1, h_2$ are assumed to  be  independent from the hash function used for the fingerprint. 

\begin{table}
\caption{Notation\label{table:notation}
}
\centering
\begin{tabular}{cp{10cm}}\toprule
$U$ & universe of all possible elements (e.g., all strings) \\
 $S$ & a set of elements from universe $U$ (also called ``keys'') \\
  $|S|$ & cardinality of the set $S$ \\
  $B$ & array of $k$-bit values \\
  $c=|B|$ & size (or capacity) of the array $B$, we set $c = \lfloor 1.23 \cdot |S| \rfloor + 32$  \\ 
  $\fingerprint$ & random hash function mapping elements of $U$ to $k$-bit values (integers in  $[0,2^k)$) \\
 $h_0, h_1, h_2$ & hash functions from $U$ to  integers in $[0,\lfloor c/3\rfloor)$,  $[\lfloor c/3\rfloor,\lfloor 2c/3\rfloor)$, $[\lfloor 2c/3 \rfloor,c)$
 respectively  \\
 $x \xor y $ & bitwise exclusive-or between two values \\
 $B[i]$ & the $k$-bit values at index $i$ (indexes start at zero) \\
$\epsilon$ & false-positive probability \\
\bottomrule
\end{tabular}
\end{table}


\subsection{Membership Tests}
The membership-test function (Algorithm~\ref{alg:lookup}) calculates the hash functions $h_0, h_1, h_2$, then constructs the expected fingerprint from those entries in table $B$, and compares it against the fingerprint of the given key. If the key is in the set, the table  contains the fingerprint and so it matches. 

The processing time includes the computation of three hash functions as well as three random memory accesses. Though other related
data structures may need fewer memory accesses, 
most modern processors can issue more than three memory accesses concurrently thanks to memory-level parallelism~\cite{jonathan2018exploiting, psaropoulos2017interleaving,akram2016boosting}. Hence, we should not expect the processing time to increase directly with the number of memory accesses. 

\begin{algorithm}
\caption{Membership test\label{alg:lookup}: returns true if the key $x$ is likely in $S$, false otherwise}
\begin{algorithmic} 
    \REQUIRE key $x \in U$
    \STATE \textbf{return} $\fingerprint(x) = B[h_0(x)] \xor B[h_1(x)] \xor B[h_2(x)]$
\end{algorithmic}
\end{algorithm}

\subsection{Construction}



The construction follows the algorithm from Botelho et al.~\cite{Botelho:2007:SSM:2394893.2394911} to build acyclic 3-partite random hypergraphs. We apply Algorithm~\ref{alg:construct} which calls Algorithm~\ref{alg:mapping} one or more times until it succeeds,
passing randomly chosen hash functions $h_0, h_1, h_2$ with each call. 
In practice, we pick hash functions by generating
a new pseudo-random \emph{seed}. 
Finally, we apply Algorithm~\ref{alg:assigning}.

\begin{algorithm}
  \caption{Construction\label{alg:construct}}
  \begin{algorithmic}
      \REQUIRE set of keys $S$
      \REQUIRE a $\fingerprint$ function
      \REPEAT
          \STATE pick three hash functions $h_0, h_1, h_2$ at random, independently from the $\fingerprint$ function
      \UNTIL{$\map(S, h_0, h_1, h_2)$ returns success with a stack  $\stack{}$ (see Algorithm~\ref{alg:mapping})}
      \STATE $B \leftarrow$ an array of size $\lfloor 1.23 \cdot |S| \rfloor + 32$ containing $k$-bit values (uninitialized)
      \STATE $\assign(\stack{}, B, h_0, h_1, h_2)$ (see Algorithm~\ref{alg:assigning})
      \STATE \textbf{return} the array $B$ and the hash functions  $h_0, h_1, h_2$
  
  \end{algorithmic}
  \end{algorithm}

Algorithm~\ref{alg:mapping} works as follows. We initialize a (temporary) array $H$ of sets of keys of size  $\lfloor 1.23 \cdot |S| \rfloor + 32$.
At the beginning, all sets are empty. Then we take each key $x$ from the set $S$, and we hash it three times ($h_0(x), h_1(x), h_2(x)$).
We append the key $x$ to the three sets indicated by the three hash values (sets $H[h_0(x)], H[h_1(x)], H[h_2(x)]$).
Most sets in the table $H$  contain multiple keys, but almost surely some  contain exactly one key. 
We keep track of the sets containing just one key.
Repeatedly, we pick one such location, append it to the output stack together with the key $x$ it contains; 
each time we remove the key  $x$ from its three locations ($h_0(x), h_1(x), h_2(x)$).
The process either terminates with a stack containing all of the keys in which case we have a success, or with a failure.

\begin{figure}\centering
\begin{tikzpicture}
\begin{axis}[
      xlabel={set size, $|S|$},
      ylabel={mapping step success probability},
      xmode=log,
      ymin = 0.4,
      legend pos=south east
      ]
\addplot[mark=triangle*, color=red] table {
3 0.986
9 0.949
27 0.865
81 0.853
243 0.767
729 0.733
2187 0.712
6561 0.815
19683 0.926
59049 0.997
177147 1.0
531441 1.0
};
\addlegendentry{$|B| = 1.23 \cdot |S| + 16$}
\addplot[mark=*, color=blue] table {
3 0.994
9 0.986
27 0.965
81 0.944
243 0.959
729 0.9
2187 0.848
6561 0.862
19683 0.948
59049 0.999
177147 1.0
531441 1.0
};
\addlegendentry{$|B| = 1.23 \cdot |S| + 32$}
\addplot[mark=square*, color=green] table {
3 1.0
9 0.998
27 0.989
81 0.979
243 0.983
729 0.986
2187 0.978
6561 0.953
19683 0.975
59049 0.999
177147 1.0
531441 1.0
};
\addlegendentry{$|B| = 1.23 \cdot |S| + 64$}
\end{axis}
\end{tikzpicture}
  \caption{Probability of mapping step, found experimentally with 1000 randomly generated sets.}
  \label{fig:probMappingStep}
\end{figure}
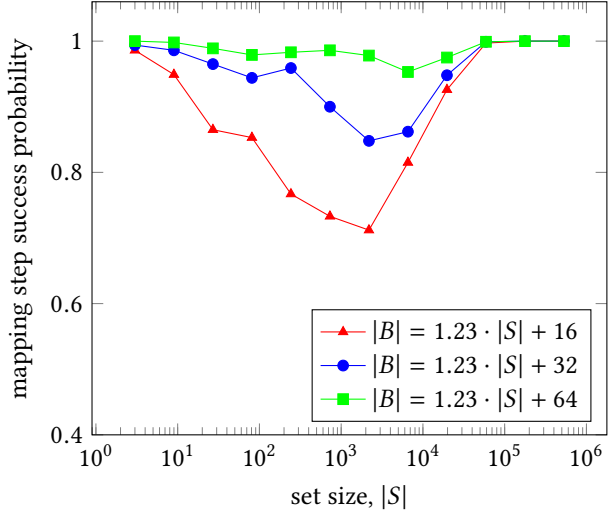

The probability of success approaches 100\% if the set is large~\cite{Molloy:2005:CRH:1072778.1072783}.
For sets of size $10^7$,  Botelho et al.~\cite{Botelho:2007:SSM:2394893.2394911} found that the probability is almost 1.  For smaller sets, we experimentally found  that the estimated probability is always greater than $0.8$ with $c = 1.23 \cdot |S| + 32$, as shown in Fig.~\ref{fig:probMappingStep}.

Algorithm~\ref{alg:mapping} runs in linear time with respect to the size of the input set $S$ as long as adding and removing
a key $x$ from a set in $H$ is done in constant time. Indeed, each key $x$ of $S$ is initially added to three sets in $H$
and removed at most once from the same three sets.

In practice, if the keys in $S$ are integer values or other fixed-length objects, we can implement the  sets  using 
an integer-value counter and a fixed-length mask  (both initialized with zeros). When adding a key, we increment the counter and
compute the exclusive-or of the key with the mask, storing the result as the new mask. We similarly remove a key by decrementing
the counter and computing the same exclusive-or. 
Even when the set is made of large or variable-length elements, it may still be practical to represent them as small fixed-length (e.g., 64-bit or 128-bit) integers by hashing: it only comes at the cost of introducing a small error when two hash values collide, an improbable event that may only minutely increase the false-probability probability.

\begin{algorithm}
\caption{Mapping Step\label{alg:mapping} ($\map$)}
\begin{algorithmic}
    \REQUIRE set of keys $S$, $k$-bit integer-valued hash functions $h_0, h_1, h_2$.
    \STATE let $c \leftarrow \lfloor 1.23 \cdot |S| \rfloor + 32$
    \STATE $H \leftarrow$ an array of size $c$ containing a set of keys (values from $S$), initially empty
    
    \FORALL{$x$ in $S$}
            \STATE append $x$ to $H[h_{0}(x)]$
            \STATE append $x$  to $H[h_{1}(x) ]$
            \STATE append $x$ to $H[h_{2}(x)]$
    \ENDFOR
    \STATE $Q\leftarrow$ initially empty queue
    \FOR{$i=0$ \textbf{to} $|H|$}
        \STATE \textbf{if} the set $H[i]$ contains a single key \textbf{then} add $i$ to $Q$ \textbf{endif}
    \ENDFOR
    \STATE $\stack{} \leftarrow$ initially empty stack
    \WHILE{queue $Q$ is not empty}
        \STATE remove an element $i$ from the queue $Q$
        \IF {the set $H[i]$ contains a single key}
            \STATE let $x$ be the sole value in the set $H[i]$
            \STATE push the pair $(x,i)$ on the stack $\stack{}$
            \FOR{$j=0$ \TO 2}
                \STATE remove $x$ from the set $H[h_{j}(x)]$
                \STATE \textbf{if} the set $H[h_{j}(x)]$ contains a single key  \textbf{then} add $h_{j}(x)$ to $Q$ \textbf{endif}
            \ENDFOR
        \ENDIF
    \ENDWHILE
    \STATE \textbf{return} success and the stack $\stack{}$ \textbf{if} $|\stack{}| = |S|$, \textbf{else} return failure
\end{algorithmic}
\end{algorithm}

We find it interesting to consider the second part of 
Algorithm~\ref{alg:mapping} when it succeeds. We iteratively 
empty the queue $Q$, one element at a time.
At iteration $t$, we add the key $x$ and the corresponding 
index $i$  to the stack if $x$ is the single key of set $H[i]$, and
we remove the key $x$ from the sets at 
locations $h_0(x), h_1(x), h_2(x)$.
Hence,  by construction, each time Algorithm~\ref{alg:mapping} 
adds a key $x$ and an index $i$ to the stack, the index $i$ is 
different from indexes $h_0(x'), h_1(x'), h_2(x')$ for 
all keys $x'$ encountered later (at time $t'>t$).

To construct the xor filter, we allocate an array $B$ large enough to store $\lfloor 1.23 \cdot |S| \rfloor + 32$ fingerprints.
We iterate over the keys and their indexes in the reverse order, compared to how they were identified  in the ``Mapping Step'' (Algorithm~\ref{alg:mapping}). 
For each key, there are three corresponding locations $h_0(x), h_1(x), h_2(x)$ in the table $B$; the index associated with the key
is one of $h_0(x), h_1(x), h_2(x)$. We set the value of $B[i]$ so that 
$B[h_0(x)] \xor B[h_1(x)] \xor B[h_2(x)] = \fingerprint(x)$.
We repeat this for each key. Each key is processed once.

By our construction, an entry in $B$ is modified at most once. After we modify an entry  
$B[i]$, then none of the values $B[h_0(x)]$, $B[h_1(x)]$, $B[h_2(x)]$ will ever be modified again.
This follows by our argument where we work through  Algorithm~\ref{alg:mapping} in reverse: 
$i$ is different from $h_0(x'), h_1(x'), h_2(x')$ for all keys $x'$ encountered so far. Remember that we use a stack, so the last entry added to the stack in Algorithm~\ref{alg:mapping} is removed first in Algorithm~\ref{alg:assigning}.
Thus, our construction is correct: we have that
\begin{align*}B[h_0(x)] \xor B[h_1(x)] \xor B[h_2(x)] = \fingerprint(x)\end{align*} for all keys $x$ in $S$ at the end of Algorithm~\ref{alg:assigning}.

\begin{algorithm}
\caption{Assigning Step\label{alg:assigning} ($\assign$)}
\begin{algorithmic}
    \REQUIRE $\stack{}$, target array for fingerprint data $B$, hash functions $h_0, h_1, h_2$
    \FOR{$(x,i)$ in  stack $\stack{}$}
    \STATE $B[i] \leftarrow 0$
    \STATE $B[i] \leftarrow \fingerprint(x) \xor B[h_0(x)] \xor B[h_1(x)] \xor B[h_2(x)]$
    \ENDFOR
\end{algorithmic}
\end{algorithm}



\subsection{Space Optimization: Xor+ Filter}
\label{sec:xorplus}

About 19\% of the entries in table $B$ are empty: for each 100~keys, we need 123~entries, and 23~are empty. 
For transmission, much of this empty space can be saved as follows:
before sending $B$, send a bit array that contains '0' for empty entries
and '1' for occupied entries. Then we only send the data of the occupied entries.
If we use $k=8$~bits, the regular xor filter needs $8 \times 1.23 = 9.84$~bits per entry, 
which we can compress in this way to $ 8 + 1.23 = 9.23$~bits per entry. 
If space usage at runtime is more important than query speed, compression can be used at runtime. 
We can get a constant time access using a rank data structure such as Rank9~\cite{vigna2008broadword}, at the expense of a  small storage overhead ($\approx$25\%), or poppy~\cite{zhou2013space} for an even smaller overhead ($\approx$3\%) at the expense of some speed. 

By changing the construction algorithm slightly, we can move most of the empty entries to the last third of the table $B$. To do so, we change the mapping algorithm so that three queues are used instead of one: one for each hash function---each hash function represents a third of the table $B$. We then process entries of the first two queues until those are empty, before we process entries from the third queue. 
Experimentally, we find that 36\% of the entries in the last third of  table $B$ are empty on average. If the rank data structure is then only constructed for this part of the table, space can be saved without affecting the membership-test performance as much, as only one rank operation is needed. 
We refer to this algorithm as ``xor+ filter'', using Rank9 as the default rank data structure. 
With the fingerprint size in bits $k$, it needs 
$k \times 1.23 \times 2/3$ bits per key for the first two thirds of the table $B$, 
$k \times 1.23 \times 1/3 \times (1-0.36)$ for the last third, plus $1.23 \times 1/3 \times 1.25$ for the Rank9 data structure.
In summary, xor+ filters use $1.0824 k + 0.5125 $~bits per entry as opposed to $1.23 k$~bit per entry for xor filters.





\subsection{Space Comparison}

\begin{figure}
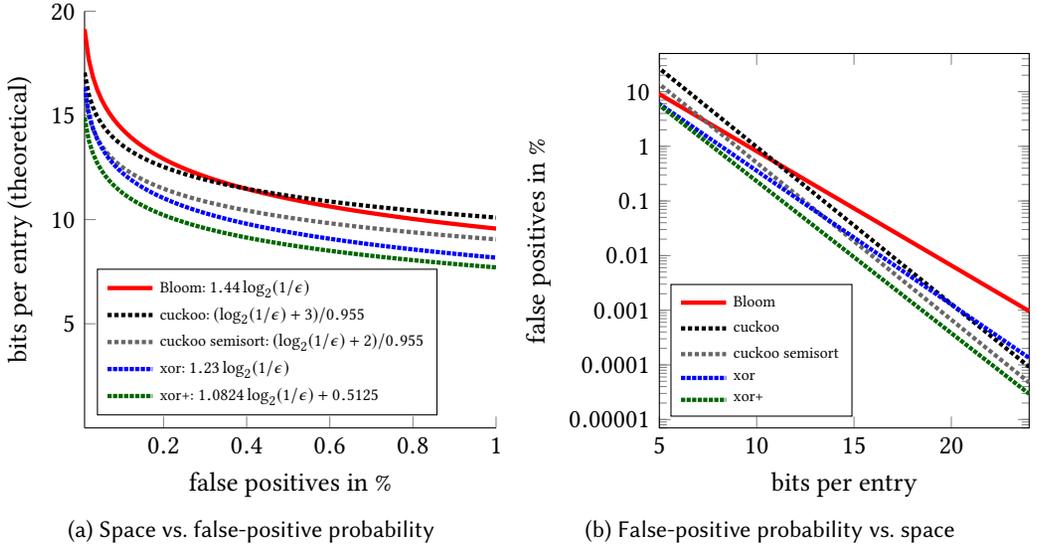

\subfloat[Space vs. false-positive probability]{
    \includegraphics[width=0.48\textwidth]{spaceusage_b.tikz}}
\subfloat[False-positive probability vs. space]{
    \includegraphics[width=0.48\textwidth]{ffpvsspace_b.tikz}}
\caption{\label{fig:spaceusagetheory_b} Theoretical memory usage for Bloom filters (optimized for space), cuckoo filter (at max. capacity) and xor filters given a desired bound on the false-positive probability. }
\end{figure}

We compare the space usage of some of the most important filters in Fig.~\ref{fig:spaceusagetheory_b}.
Bloom filters are more space efficient than cuckoo filters at a false-positive probability of 0.4\% or higher.

For very low false-positive probabilities ($5.6 \times 10^{-6}$), cuckoo filters at full capacity use less space than xor filters. However, we are not aware of any system that uses such a low false-positive probability: most systems seem to use between 8 and 20~bits per key~\cite{Sears:2012:BGP:2213836.2213862,dharmapurikar2003deep}. Thus we expect xor and xor+ filters to use less memory in practice.

\section{Experiments}

We follow  Fan et al.'s testing procedure~\cite{Fan:2014:CFP:2674005.2674994}; we started from their software project~\cite{githubCuckoofilter}. 
Like them,  we use 64-bit keys as set elements. 
We build a filter based on a set of 10M or 100M keys.
We build a distinct set made of 10M queried keys. This set of queried keys is created by mixing some of
the keys from the original set, and some keys not present in the original set. We use different fractions (e.g., 0\%, 25\%, 50\%, 75\% and 100\%) of the keys  in the original set. The benchmark  counts the number of queried keys that are possibly in the set according to the filter. 
The benchmark is single threaded and calls the membership-test functions with different keys in a loop. We disable inlining of the functions to prevent compilers from 
unduly optimizing the benchmark which counts the number of matching keys. 

We run benchmarks on Intel processors with Skylake microarchitecture: 
an Intel i7-6700 processor running at 3.4\,GHz, with 8\,MB of L3 cache.
The software is compiled with the GNU~GCC~8.1.0 compiler to a 64-bit Linux executable with the flags \texttt{-O3 -march=native}.
For each filter, we run 3~tests, and report the median. 
Our error margin is less than 3\%.
The C++ source code of the filter implementations
and the benchmark is available\footnote{\url{https://github.com/FastFilter/fastfilter_cpp} (release 1.0), see ``Benchmarking'' section.}.
For some algorithms including all the xor and xor+ filters, we have also implemented Java versions\footnote{\url{https://github.com/FastFilter/fastfilter_java}} and well as a Go version\footnote{\url{https://github.com/FastFilter/xorfilter}}  and a pure C version\footnote{\url{https://github.com/FastFilter/xor_singleheader}},
but the benchmarks are using C++.

For all implementations, we use a randomly seeded Murmur finalizer~\cite{ivanchykhin2017regular} to compute the fingerprint from the key, as described in Algorithm~\ref{alg:hash}. We choose this option instead of faster alternatives so that even non-random keys work well and do not result in higher-than-expected false-positive probabilities, or construction failure  in the case of the cuckoo filter. For our tests, we use pseudo-randomly generated keys; we also tested with sequentially generated keys and found no statistically significant difference compared to using random keys after introducing the Murmur finalizer.

All implementations need to reduce a hash value $x$ to the range $\{0,\ldots,m - 1\}$ where $m$ is not necessarily a power of two.
Where this is needed,  we do not use the relatively slow modulo operation $x \mod m$ for performance reasons.
Instead, starting with  32-bit values $x$ and $m$ and computing their full 64-bit product $x \times m$, we use the faster multiply-shift combination $(x \times m) \div {2^{32}} = (x \times m) \mathrm{~>>~} 32$~\cite{Lemire:2019:FRI:3309768.3230636}.

\begin{algorithm}
\caption{64-bit hash function\label{alg:hash}}
\begin{algorithmic}
    \REQUIRE key $x$, seed $s$
    \STATE $h \leftarrow x + s$
    \STATE $h \leftarrow (h \xor (h >> 33)) * \mathrm{0xff51afd7ed558ccd}$
    \STATE $h \leftarrow (h \xor (h >> 33)) * \mathrm{0xc4ceb9fe1a85ec53}$
    \STATE \textbf{return} $h \xor (h >> 33)$
\end{algorithmic}
\end{algorithm}


\subsection{Filter Implementations}

We run tests against the following filters:

\begin{itemize}

\item Bloom filter: We implemented the standard Bloom filter algorithm with configurable  false-positive probability (FPP) and size. We test with 8, 12, and 16 bits per key, and the respective number of hash functions $k$ that are needed for the lowest false-positive probability. For fast construction and membership test, we hash only once with a 64-bit function, treated as two 32-bit values $h_1(k)$ and $h_2(k)$. The Bloom filter hash functions are $g_i(k) = h_1(k) +i \cdot h_2(k)$ for $i=0,\ldots,k-1$. 

\item Blocked Bloom filter: We use a highly optimized blocked Bloom filter from Apache Impala\footnote{\url{https://impala.apache.org}}, which is also used in the cuckoo filter software project~\cite{githubCuckoofilter}. We modified it so the size is flexible and not restricted to $2^n$. It is designed for Intel AVX2 256-bit  operations; it is written using low-level Intel intrinsic functions. The advantage of this algorithm is the membership-test speed: each membership test is resolved from one cache line only using few instructions. The main disadvantage is that it is larger than regular Bloom filters.

\item Cuckoo filter (C): We started with the cuckoo filter implementation from the original authors~\cite{githubCuckoofilter}. 
We reduce the maximum load from 0.96 to 0.94, as otherwise construction occasionally fails. The reduced maximum load is apparently the recommended workaround suggested by the cuckoo filter authors. Though it is outside our scope to evaluate whether it is always a reliable  fix, it was sufficient in our case. This reduction of the maximum load slightly worsens ($\approx $2\%) the memory usage of cuckoo filters. 
In the original reference implementation~\cite{Fan:2014:CFP:2674005.2674994}, the size of the filter is restricted to be a power of two, which means up to 50\% of the space is unused. Wasting so much space seems problematic, especially since it does not improve the false-positive probability. Therefore, we modified it so the size is flexible and not restricted to $2^n$. This required us to slightly change the calculation for the alternate location $l_2(x)$ for a key $x$ from the first location $l_1(x)$ and the fingerprint $f(x)$. Instead of $l_2(x) = l_1(x) \xor h(f(x))$ as in Fan et al.~\cite{Fan:2014:CFP:2674005.2674994}, we use $l_2(x) = \mathrm{bucketCount} - l_1(x) - h(f(x))$, and if the result is negative we add $\mathrm{bucketCount}$.
We use 12-bit and 16-bit fingerprints.


\item Cuckoo semi-sorted (Css): We use the semi-sorted cuckoo filter reference implementation, modified in the same way as the regular cuckoo filter. From the original Fan et al.~\cite{Fan:2014:CFP:2674005.2674994} source release, we could only get one variant to work correctly, the version with a fingerprint size of 13~bits. Other versions have a non-zero false negative probability. 

\item Golomb-compressed sequence (GCS): Our implementation uses an average bucket size of 16, and Golomb Rice coding. We use a fingerprint size of 8~bits. 

\item Xor: Our xor and xor+ filters as described in \S~\ref{xor}. We use 8-bit and 16-bit fingerprints. 

\end{itemize}


\subsection{Construction Performance}

We present the construction times for 10~million and 100~million keys in Table~\ref{table:constructionTimeNsKey}. 
All construction algorithms are single-threaded; we did not investigate multi-threaded construction. 
For reference, we also present the time needed to sort the 64-bit keys using the C++ standard sorting algorithm (\texttt{std::sort}), on the same platform.

During construction, the blocked Bloom filter is clearly the fastest data structure. For the 100~million case, the semi-sorted variant of the cuckoo filter is the slowest. Construction of the xor filter with our implementation is roughly half as fast as the cuckoo filter and the Bloom filter, which have similar performance.

\begin{table}
\caption{Construction time in nanoseconds per key, rounded to 10 nanoseconds.}\label{table:constructionTimeNsKey}
\begin{tabular}{lrr}
\toprule
algorithm & 10 million keys & 100 million keys \\
  \midrule
Blocked Bloom & 10 ns/key & 20 ns/key \\
Bloom 8 & 40 ns/key & 70 ns/key \\
Bloom 12 & 60 ns/key & 90 ns/key \\
Bloom 16 & 90 ns/key & 130 ns/key \\
Cuckoo semiSort 13 & 130 ns/key & 200 ns/key \\
Cuckoo 12 & 80 ns/key & 130 ns/key \\
Cuckoo 16 & 90 ns/key & 120 ns/key \\
GCS & 160 ns/key & 190 ns/key \\
Xor 8 & 110 ns/key & 130 ns/key \\
Xor 16 & 120 ns/key & 130 ns/key \\
Xor+ 8 & 160 ns/key & 180 ns/key \\
Xor+ 16 & 160 ns/key & 180 ns/key \\
(Sorting the keys) & 80 ns/key & 90 ns/key \\  
  \bottomrule
\end{tabular}
\end{table}

\subsection{Query Time Versus Space Overhead}

\begin{figure}
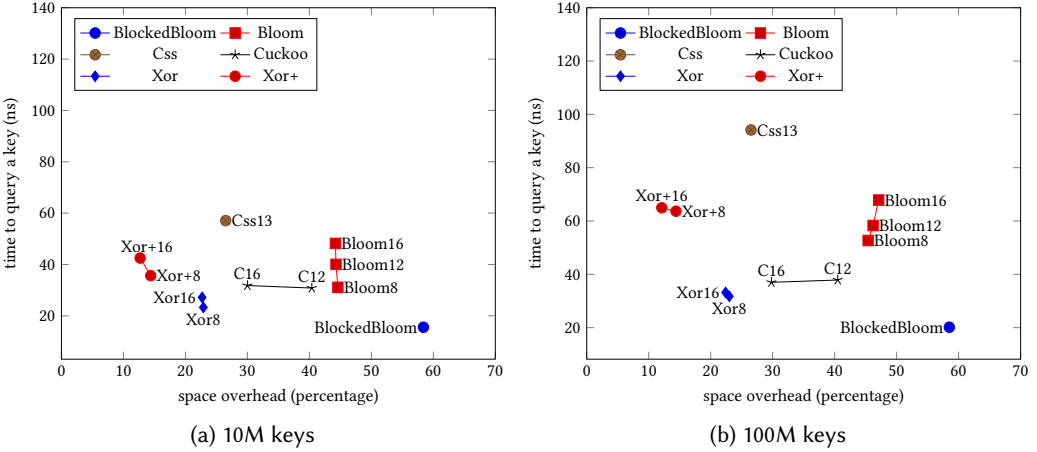

\subfloat[10M keys]{
    \includegraphics[width=0.49\textwidth]{timevsspace10M-25-find.tikz}
  }
  \subfloat[100M keys]{
  \includegraphics[width=0.49\textwidth]{timevsspace100M-25-find.tikz}
  }
  \caption{Query time vs.\ space overhead, 25\% find}
  \label{fig:qtvsspc25}
\end{figure}

\begin{figure}
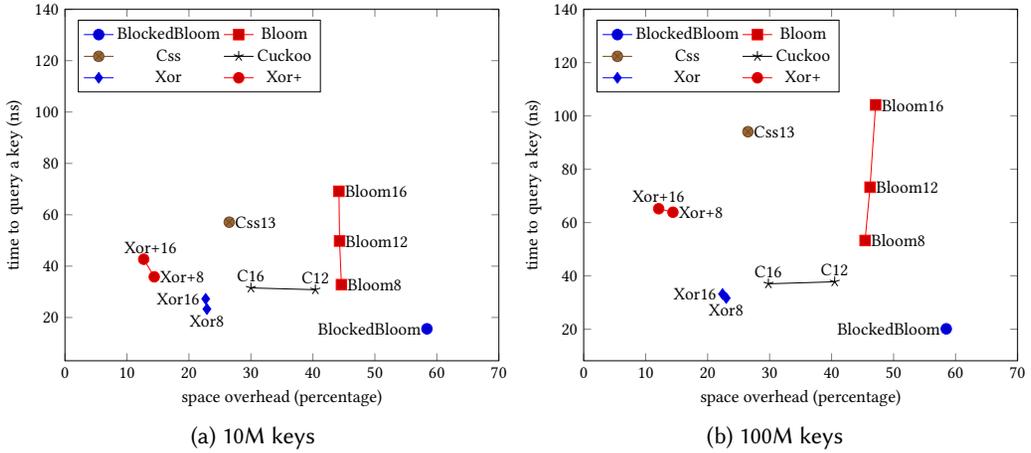

\subfloat[10M keys]{
    \includegraphics[width=0.49\textwidth]{timevsspace10M.tikz}}
    \subfloat[100M keys]{
    \includegraphics[width=0.49\textwidth]{timevsspace100M.tikz}
}
  \caption{Query time vs.\ space overhead, 100\% find}
  \label{fig:qtvsspc100}
\end{figure}

We present the performance numbers for the case where 25\% of the searched entries are in the set in Fig.~\ref{fig:qtvsspc25},
and  in the case where all searched entries are in the set in  Fig.~\ref{fig:qtvsspc100}. 
The results are presented in tabular form in Table~\ref{table:lookupBenchmarkResults}, where we include the Golomb-compressed sequence.

\begin{table}
\caption{Membership-test benchmark results, 25\% find. Timings are in nanosecond per query. \label{table:lookupBenchmarkResults}}
\subfloat[10M keys]{
\begin{tabular}{lrrr}
\toprule
Name & Time (ns) & Bits/key & FPP  \\
\midrule
Blocked Bloom & 16 & 10.7 & 0.939  \\
Bloom 8 & 31 & 8.0 & 2.161  \\
Bloom 12 & 40 & 12.0 & 0.313 \\
Bloom 16 & 48 & 16.0 & 0.046 \\
Cuckoo semiSort 13 & 57 & 12.8 & 0.092  \\
Cuckoo 12 & 31 & 12.8 & 0.183  \\
Cuckoo 16 & 32 & 17.0 & 0.012 \\
GCS & 137 & 10.0 & 0.389  \\
Xor 8 & 23 & 9.8 & 0.389  \\
Xor 16 & 27 & 19.7 & 0.002 \\
Xor+ 8 & 36 & 9.2 & 0.390  \\
Xor+ 16 & 43 & 17.8 & 0.002  \\
\bottomrule
\end{tabular}
}
\subfloat[100M keys]{
\begin{tabular}{lrrrr}
\toprule
 & Time (ns) & Bits/key & FPP \\
\midrule
 & 20 & 10.7 & 0.941 \\
 & 53 & 8.0 & 2.205  \\
& 58 & 12.0 & 0.339  \\
 & 68 & 16.0 & 0.053  \\
  & 94 & 12.8 & 0.092 \\
 & 38 & 12.8 & 0.184  \\
& 37 & 17.0 & 0.011  \\
 & 220 & 10.0 & 0.390  \\
 & 32 & 9.8 & 0.391  \\
 & 33 & 19.7 & 0.001  \\
 & 64 & 9.2 & 0.389  \\
 & 65 & 17.8 & 0.002 \\
\bottomrule
\end{tabular}
}

\end{table}

Unlike xor and cuckoo filters, the Bloom filter membership-test timings are sensitive to the fraction of keys present in the set. When an entry is not in the set, only a few bits need to be accessed, until the query function finds an unset bit and returns.
The Bloom filter is slower if an entry exists in the set, as it has to check all bits; this is especially the case for low false-positive probabilities. See Fig.~\ref{fig:qtvsspc100}. 



\begin{figure}
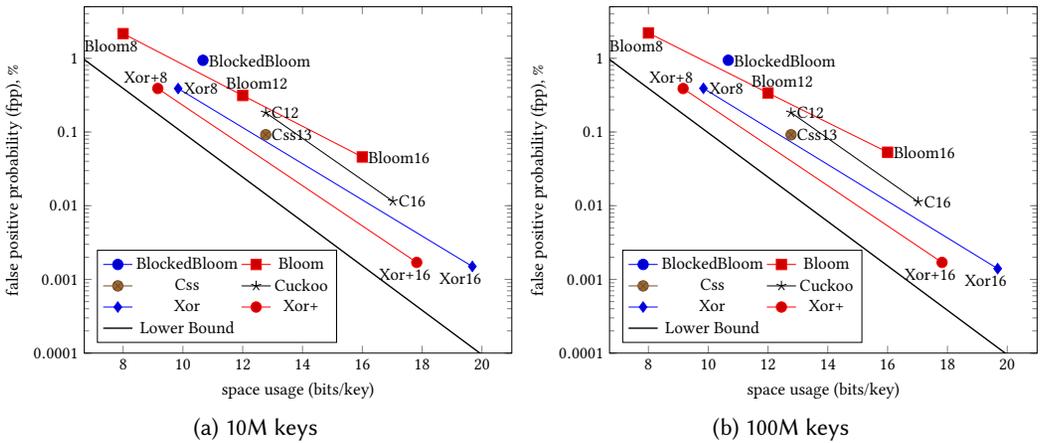

\subfloat[10M keys]{
\includegraphics[width=0.49\textwidth]{fppvsbitslog10.tikz}
}
  \subfloat[100M keys]{
    \includegraphics[width=0.49\textwidth]{fppvsbitslog100.tikz}
}
\caption{FPP vs.\ space usage in bits/key, log scale FPP\label{fig:fppvsspace}}
\end{figure}

Ignoring query time,
Fig.~\ref{fig:fppvsspace} shows that Cuckoo~12 (C12) has memory usage that is close to Bloom filters. The cuckoo filter only uses much less space than Bloom filters for false-positive probabilities well below 1\% (Cuckoo~16 or C16).
In our experiments, the cuckoo filter,
and the slower semi-sorted cuckoo filter (Css), always use more space than the xor filter.
These experimental results match the theoretical results presented in Fig.~\ref{fig:spaceusagetheory_b}.


The xor filter provides good query-time performance while using little space, even for moderate false-positive probabilities.

\subsection{Discussion}

We attribute the  good membership-test performance of xor filters mainly to the following reasons.
Xor filters use exactly 3~memory accesses, independent of the false-positive probability. These memory accesses
can be executed in parallel by the memory subsystem. The number of instructions meanwhile is small and
there are no branches.


For a false-positive probability of 1\%, the standard Bloom filter needs more memory accesses for a match, 
and even more so for lower false-positive probabilities. 
The Bloom filter uses between 41~and 105~instructions per key, depending on the number of set bits set and false-positive probability.
For a miss (if the key is not in the set), on average fewer memory accesses are needed, but there might be mispredicted  branches with accompanying penalties.

The cuckoo filter uses exactly 2~memory accesses, and 66 to 68 instructions per key (depending on fingerprint size).
The xor filter uses exactly 3~memory accesses, but only about 48~instructions per key.
Processors execute complex machine instructions using low-level instructions called $\mu$ops.
A processor like our Skylake can support
up to 10~outstanding memory requests per core, limited by an instruction  reorder buffer of 200~$\mu$ops. In the absence of mispredicted branches and long dependency chains,  the capacity of the instruction buffer becomes a limitation~\cite{beamer2015locality}.
It is likely the reason why the cuckoo filter and the xor filter have similar membership-test performance.  That is, while the cuckoo filter has fewer memory accesses, it generates more instructions which makes it harder for the processor to fetch as many memory requests as it could.

In our benchmarks, the blocked Bloom filter is the only algorithm that is clearly faster than the xor filter. This is most likely due to only having one memory access, and highly optimized code, using SIMD instructions specific to recent x64 processors. It needs fewer memory accesses and fewer instructions than its competitors.  
It might be difficult to implement a similarly efficient approach in a higher-level language like Java, or using solely portable code. If memory usage or low false-positive probability are a primary concern, the blocked Bloom filter may not be a good choice.

While an xor filter is immutable, we believe that it is not a  limitation for many important applications; competitive alternatives all have limited mutability in any case. Approximate filters that support fast mergers or additions (e.g., Bloom filters) require the original filters to have extra capacity. The update may even fail in the case of Cuckoo filters. Re-building the filter can maintain an optimal size. In multithreaded systems, immutability avoids the overhead of synchronization mechanisms to maintain concurrency.






\section{Conclusion and Future Work}

Xor filters are typically faster than Bloom filters,
and they save about 15\% in memory usage. While the construction of xor filters is  slower than Bloom filters ($\approx 2 \times $), we expect that the construction is a one-time cost amortized over many queries. 
Future work could consider batched queries~\cite{Breslow:2018:MFF:3213880.3232248} to improve performance. It might also be possible to partially parallelize the construction of the filters.

\appendix
\section{Original Versus Compact Cuckoo Filters}
In Table~\ref{table:constructionTimeNsKeyTradeOff}, we compare the 
original implementations of cuckoo filters which require that the filter
size be a power of two, with our more compact implementation. Given 10~million keys, the memory usage of cuckoo filters using the original implementation is not competitive. However, the construction time 
is reduced with an overallocated filter because hash collisions are less frequent. Similarly, the query times (25\% of the entries in the set) are about 10\% smaller in the original implementation. However, if we choose a number of keys near a power of two (e.g., 31.5~million keys), the original and compact implementations have nearly the same memory usage, construction times, and query speeds.

\begin{table}
\caption{Construction time, memory usage and query time (25\% of the entries in the set) for original and compact cuckoo filters with 10 million keys.}\label{table:constructionTimeNsKeyTradeOff}
\begin{tabular}{rrrr}
\toprule
Name & Construction Time & Memory & Query Time \\
  \midrule
Cuckoo 12 & 80 ns/key & 12.8~bits/key  & 31~ns/key\\
Cuckoo 12 (original) & 40 ns/key & 20.1~bits/key & 30~ns/key\\[0.2cm]
Cuckoo 16 & 90 ns/key & 17.0~bits/key  & 32~ns/key\\
Cuckoo 16 (original) & 40 ns/key & 26.8~bits/key & 28~ns/key\\
  \bottomrule
\end{tabular}
\end{table}

\section{Quotient  and Morton   Filters}
We consider quotient filters~\cite{Pandey:2017:GCF:3035918.3035963} (CQF) experimentally.
 We use the reference implementation~\cite{githubCQF}. The implementation
 relies on assembly code optimized for recent x64 processors.
 As with cuckoo filters, the original implementation requires that the capacity 
 be  a power of two. 
Table~\ref{table:cqflookupBenchmarkResults} shows that the query time of the reference quotient-filter implementation is several times the query time of competitive approaches like cuckoo or xor filters. We also consider Morton filters~\cite{Breslow:2018:MFF:3213880.3232248} with the reference implementation~\cite{githubMortonfilter}. Morton filters answer one-at-a-time  queries at half the speed of 8-bit xor filters, despite similar false-positive probabilities and memory usage. 

\begin{table}\centering
\caption{Performance of counting quotient filters (CQF) and Morton filters. We include Xor~8 results for comparison. For queries, we report the results corresponding to 25\% of the entries being in the set. \label{table:cqflookupBenchmarkResults} }
\begin{tabular}{lrrrr}
\toprule
Name & Query Time (ns/key) & Bits/key & FPP & volume \\
\midrule
Xor~8 & 23 & 9.8 & 0.39 & 10M \\
CQF & 64 & 17.0 & 0.23 & 10M \\
Morton & 47 & 11.7 & 0.31 & 10M \\\midrule
Xor~8 & 32 & 9.8 & 0.39 & 100M \\
CQF & 88 & 13.6 & 0.29 & 100M \\
Morton & 65 & 11.7 & 0.31 & 100M \\
\bottomrule
\end{tabular}
\end{table}

\begin{acks}
We are grateful to J.~Apple for his feedback.
\end{acks}

\bibliographystyle{ACM-Reference-Format}
\bibliography{xorfilter}

\end{document}